\documentclass[a4paper,aps,twocolumn,prl,showpacs,amsmath,amssymb,superscriptaddress,draft=false]{revtex4-1}

\usepackage{graphicx}
\usepackage{dcolumn}
\usepackage{bm}
\usepackage{longtable}
\usepackage{xcolor}
\usepackage{changes}
\usepackage[]{subfigure}

\makeatletter
\newcommand\figcaption{\def\@captype{figure}\caption}
\newcommand\tabcaption{\def\@captype{table}\caption}

\makeatother

\begin{document}


\title{Tailoring Materials for Mottronics: \\Excess Oxygen Doping of a Prototypical Mott Insulator}

\author{P. Scheiderer}
\affiliation{Physikalisches Institut and R\"ontgen Center for Complex Material Systems (RCCM), Universit\"at W\"urzburg, Am Hubland,
D-97074 W\"urzburg, Germany}
\author{M. Schmitt}
\affiliation{Physikalisches Institut and R\"ontgen Center for Complex Material Systems (RCCM), Universit\"at W\"urzburg, Am Hubland,
D-97074 W\"urzburg, Germany}
\author{J. Gabel}
\affiliation{Physikalisches Institut and R\"ontgen Center for Complex Material Systems (RCCM), Universit\"at W\"urzburg, Am Hubland,
D-97074 W\"urzburg, Germany}
\author{M. Zapf}
\affiliation{Physikalisches Institut and R\"ontgen Center for Complex Material Systems (RCCM), Universit\"at W\"urzburg, Am Hubland,
D-97074 W\"urzburg, Germany}
\author{M. St\"ubinger}
\affiliation{Physikalisches Institut and R\"ontgen Center for Complex Material Systems (RCCM), Universit\"at W\"urzburg, Am Hubland,
D-97074 W\"urzburg, Germany}
\author{\\ P. Sch\"utz}
\affiliation{Physikalisches Institut and R\"ontgen Center for Complex Material Systems (RCCM), Universit\"at W\"urzburg, Am Hubland,
D-97074 W\"urzburg, Germany}
\author{L. Dudy}
\affiliation{Physikalisches Institut and R\"ontgen Center for Complex Material Systems (RCCM), Universit\"at W\"urzburg, Am Hubland,
D-97074 W\"urzburg, Germany}
\author{C. Schlueter}
\affiliation{Diamond Light Source Ltd., Didcot, Oxfordshire OX11 0DE, United Kingdom}
\author{T.-L. Lee}
\affiliation{Diamond Light Source Ltd., Didcot, Oxfordshire OX11 0DE, United Kingdom}
\author{M. Sing}
\author{R. Claessen}
\affiliation{Physikalisches Institut and R\"ontgen Center for Complex Material Systems (RCCM), Universit\"at W\"urzburg, Am Hubland,
D-97074 W\"urzburg, Germany}

\date{\today}

\begin{abstract}
The Mott transistor is a paradigm for a new class of electronic devices---often
referred to by the term Mottronics---, which are based on charge
correlations between the electrons. Since correlation-induced insulating phases
of most oxide compounds are usually very robust, new methods have to be
developed to push such materials right to the boundary to the
metallic phase in order to enable the metal-insulator transition to be switched
by electric gating.

Here we demonstrate that thin films of the prototypical Mott insulator LaTiO$_3$ grown
by pulsed laser deposition under oxygen atmosphere are readily tuned by excess
oxygen doping across the line of the band-filling controlled Mott transition
in the electronic phase diagram. The detected insulator to metal transition is characterized by a strong change in resistivity of several orders of magnitude.
The use of suitable substrates and capping layers to inhibit oxygen diffusion facilitates full control of the oxygen content and renders the films stable against exposure to ambient conditions, making LaTiO$_{3+x}$ a
promising functional material for Mottronics devices.
\end{abstract}
\maketitle
\section*{}
{
The rich electronic phase diagram of transition metal oxides caused by strong electron correlations offers new perspectives for future device applications, often referred to as Mottronics.$^\text{\cite{takagi_emergent_2010,ahn_electric_2003}}$ One particularly striking
approach is the Mott transistor. In contrast to semiconductor devices, here an electronic phase transition rather than the mere manipulation of charge is employed as a switch.
One possible realization is a Mott insulator (OFF-state) that is reversibly driven into a metallic phase (ON-state).
Mott insulators as "failed metals" have metallic charge carrier densities at integer band-filling, but exhibit insulating behavior due to strong on-site Coulomb interactions of the valence electrons.
Detuning the band-filling, e.g., by applying an external electric field, can trigger the Mott transition (MT) into the correlated metal phase rendering all previously localized electrons mobile in the ON-state.$^\text{\cite{son_heterojunction_2011,nakano_collective_2012,inoue_taming_2008,mazza_field-driven_2016,newns_mott_1998,imada_metal-insulator_1998,inoue_electrostatic_2005}}$
Besides the large ON/OFF ratio and fast switching speed in such devices, the high charge carrier concentration allows for further miniaturization beyond the current limits set by the extremely small number of carriers in nanoscale semiconductor devices.$^\text{\cite{takagi_emergent_2010,castell_dopant_2003,inoue_electrostatic_2005,Chudnovskiy_2002,son_heterojunction_2011}}$

The downside of the high charge carrier density, however, is the need for large electric fields to significantly change the band-filling and thereby trigger the phase transition.$^\text{\cite{son_heterojunction_2011}}$
Such high fields can indeed be achieved by liquid ion gating, which was successfully applied to demonstrate the advanced
functionalities of Mott transistors in VO$_2$ and NiNdO$_3$, but may not be suitable for scalable electronics.$^\text{\cite{nakano_collective_2012,scherwitzl_electric-field_2010}}$

Recently, the realization of a Mott transistor involving only moderate electric gate fields was suggested. Dynamical mean field theory (DMFT) based simulations have demonstrated that materials in the Mott insulating phase, but already near the MT, are highly sensitive to external parameters such as strain and temperature, and also electric fields.$^\text{\cite{zhong_electronics_2015,mazza_field-driven_2016}}$
To reach the operating point close to the transition into the correlated metal phase, the development of methods to tune a material's position in the electronic phase diagram is crucial.
In principle there are two ways to approach the MT, either by decreasing the on-site Coulomb interaction relative to the electron bandwidth (bandwidth controlled MT) or by changing the band-filling (filling controlled MT).$^\text{\cite{imada_metal-insulator_1998}}$

Here, we report on the filling controlled MT in LaTiO$_{3+x}$ (LTO) thin films driven by excess oxygen doping.
The material's tendency to over-oxidize makes the fabrication of LTO thin films challenging, but offers a precise tunability of the band-filling once full control over the oxygen stoichiometry is gained.
\begin{figure*}[ht]
{
\includegraphics[width = 1\textwidth]{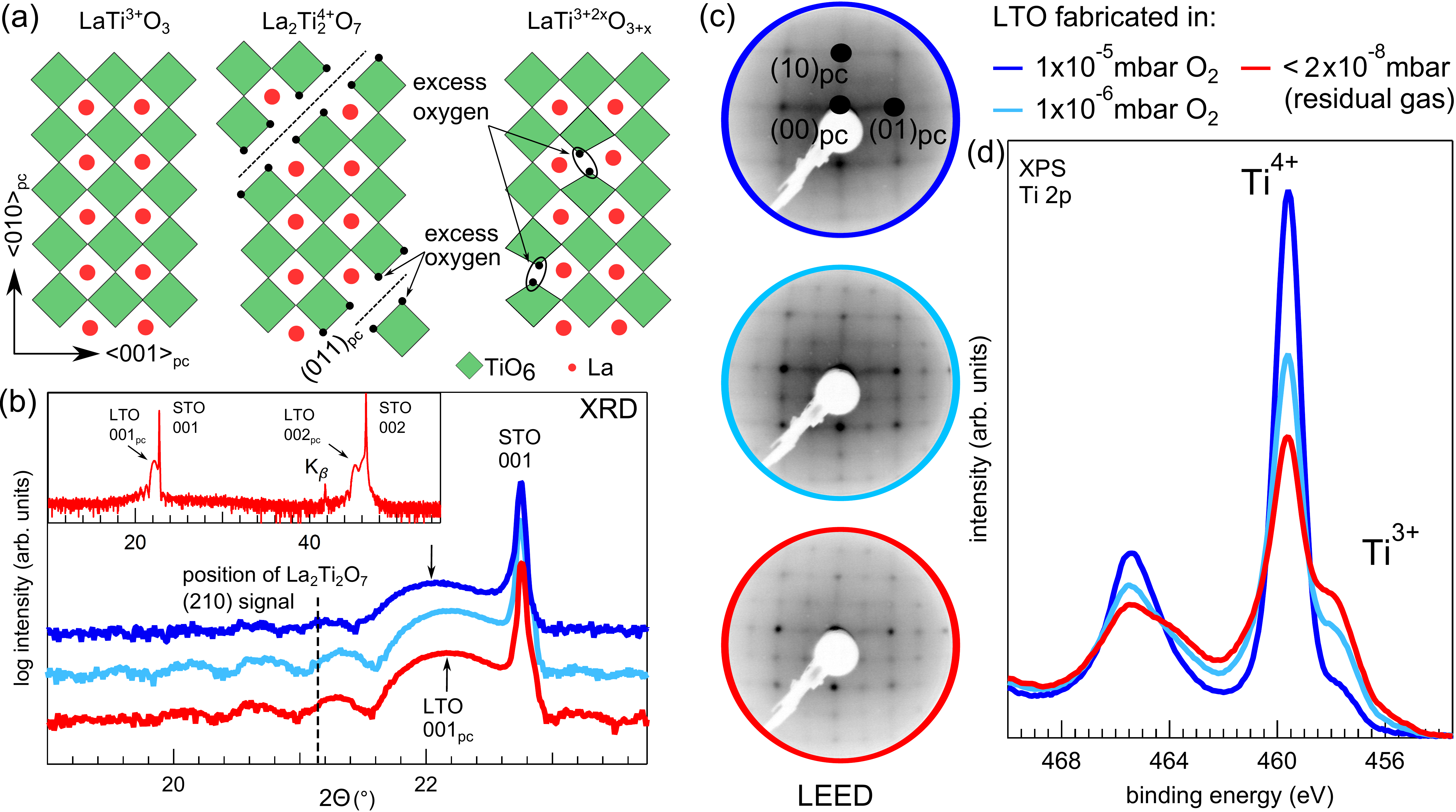}
}
\caption{
\label{Fig1}
(a) Crystal structures of LaTiO$_3$, La$_2$Ti$_2$O$_7$, and LaTiO$_{3+x}$ (for details see text). 
(b) XRD patterns of  LTO films grown by PLD on STO (001) substrates at various oxygen growth pressures. Phase-purity is established by the absence of Bragg peaks for the oxygen-rich La$_2$Ti$_2$O$_7$ phase or other Ti oxides (see also inset with larger scattering angle range). K$_{\beta}$ denotes the satellite line caused by the Cu K$_{\beta}$ part of the x-ray source.
(c) LEED patterns reflect the structural quality of the film surface, which improves as the oxygen growth pressure is reduced.
(d) Ti 2$p$ spectra probed by in situ XPS reveal a strong over-oxidation towards Ti$^{4+}$.
}
\end{figure*}

Due to their strong on-site Coulomb interactions and narrow bandwidths 3 $d^1$ perovskites are intensively studied as prototypical Mott systems. The series SrVO$_3$-CaVO$_3$-LaTiO$_3$-YTiO$_3$ exhibits a trend towards increasing electron localization from an enhanced effective electron mass in the metallic vanadates to increasing charge gaps (0.2...1$\;$eV) in the Mott insulating titanates.$^\text{\cite{fujimori_evolution_1992,arima_variation_1993}}$
Recently, it was pointed out that the drastic change of the electronic properties in these seemingly similar materials is based on the progression of an orthorhombic distortion towards the GdFeO$_3$-type structure. In the undistorted cubic perovskite SrVO$_3$ the three $t_{2g}$ bands are degenerate with an effective band-filling of $\frac{1}{6}$ favoring the correlated metal phase. In the distorted titanates, however, the splitting of the $t_{2g}$ levels gives rise to Mott insulating behavior.$^\text{\cite{haverkort_determination_2005,pavarini_mott_2004}}$ Stoichiometric LaTiO$_3$ is therefore a prototypical single band Mott insulator and, as indicated by its small Mott gap, already close to the transition into the correlated metal phase.

\begin{figure*}[htbp]
{
\includegraphics[width = 1\textwidth]{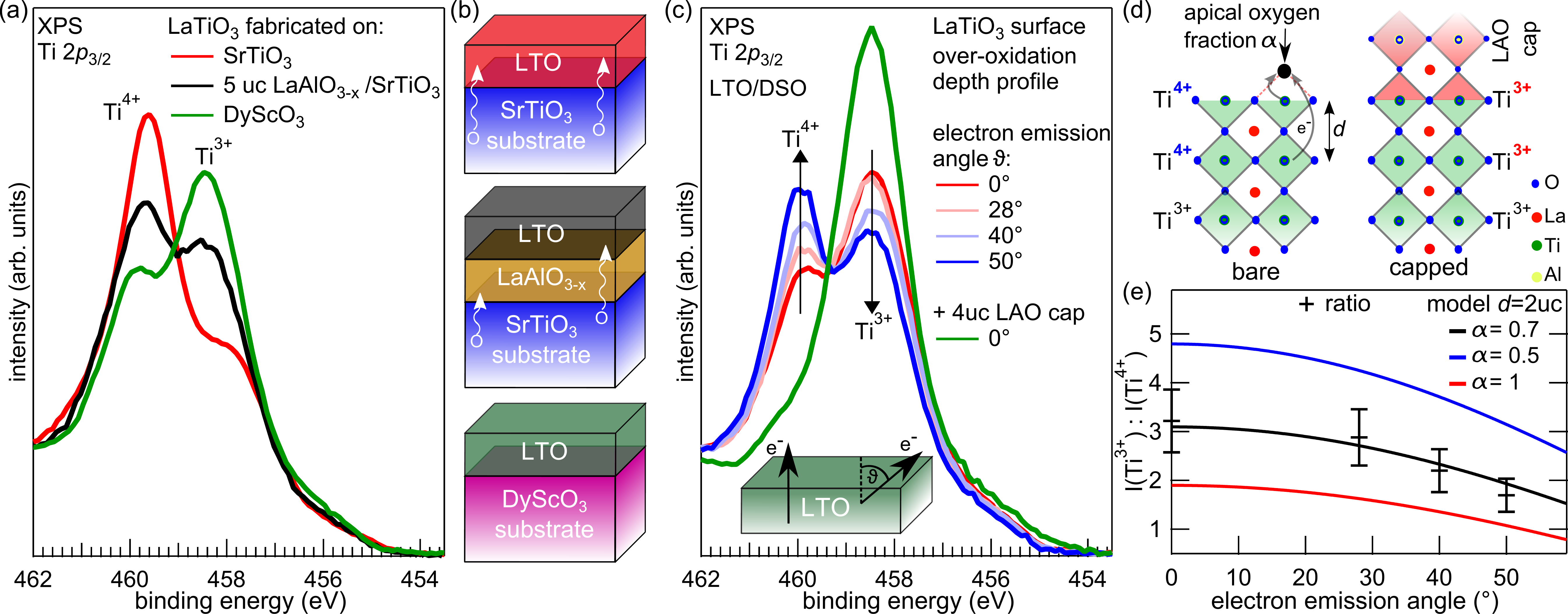}
}
\caption{
\label{Fig2}
(a) Ti 2$p_{3/2}$ XPS spectra of LTO films prepared on various substrates. A strong impact of the substrate choice on the titanium valence is detected. (b) Oxygen out-diffusion from the different substrates into the LTO films (see text for details). (c) Angle-dependent XPS spectra of the Ti 2$p_{3/2}$ line measured on the bare LTO film surface together with data from an LAO-capped LTO film, both grown on a DSO substrate.
(d) Microscopic model describing the impact of additional oxygen adsorbed at apical surface sites on the electronic structure of the LTO top layers. (e) Spectral weight ratio of the tri- to tetravalent titanium components as a function of electron emission angle, obtained from the spectra measured on the bare LTO surface presented in (c). A quantitative evaluation based on the microscopic model yields a coverage with apical oxygen of $\alpha \approx$ 75\% and an extension $d$ of the Ti$^{4+}$-containing surface layer of 2 uc.
}
\end{figure*}
Further approaching the MT appears possible by $p$-doping with excess oxygen, since LaTiO$_3$ is the $n=\infty$ endmember of the homologous series La$_n$Ti$_n$O$_{3n+2}$.$^\text{\cite{lichtenberg_synthesis_2001}}$ Here, excess oxygen is hosted in the (011)-planes of the pseudocubic (pc) perovskite unit cell with the index $n$ denoting the distance between these planes in unit cells ranging from $n=\infty$ to $n=4$. The structure of the two endmembers LaTiO$_3$ and La$_2$Ti$_2$O$_7$ with a corresponding configuration of $d^1$ (Ti$^{3+}$) and $d^0$ (Ti$^{4+}$) is sketched in Figure \ref{Fig1}(a).
The existence of these oxygen rich phases poses challenges for the fabrication of stoichiometric LaTiO$_3$ thin films, since the energetically favored valence is Ti$^{4+}$ and unintentional and uncontrolled over-oxidation may easily occur.

In order to systematically investigate the effects of over-oxidation we fabricated a series of LTO thin films by pulsed laser deposition (PLD) at typical growth parameters$^\text{\cite{ohtomo_epitaxial_2002,disa_orbital_2015,ohtomo_artificial_2002,biscaras_two-dimensional_2012,wong_metallicity_2010,takizawa_photoemission_2006,chang_layer-by-layer_2013,lin_final-state_2015}}$ on SrTiO$_3$ (STO) substrates in oxygen background pressures of $10^{-5}\;$mbar and lower. The structural characterization by x-ray diffraction (XRD) and low-energy
electron diffraction (LEED) is depicted in Figure \ref{Fig1}(b) and (c) and reveals a general trend: the quality of the diffraction pattern improves for films fabricated in lower oxygen pressure as judged by the Laue intensity oscillations in XRD and the overall contrast and width of the Bragg peaks in LEED. Furthermore, an increase of the lattice constant is detected by XRD for growth in an oxygen pressure of $10^{-5}\;$mbar, as indicated by arrows in Figure \ref{Fig1}(b). Besides LaTiO$_3$ no other phases are detected. Especially the absence of the La$_2$Ti$_2$O$_7$ signal at 2$\Theta=21.15^{\circ}$ and any diffraction peaks from binary titanate compounds as seen in Refs.\onlinecite{ohtomo_epitaxial_2002} and \onlinecite{grisolia_structural_2014} demonstrate the phase purity of the films. Note that the 2x2 surface reconstruction with respect to the pc unit cell in LEED is in line with the GdFeO$_3$-type perovskite structure.

The effect of doping is studied by monitoring the valence of the titanium cations as probed by x-ray photoelectron spectroscopy (XPS) of the Ti 2$p$ core level.
Due to the screening by the additional 3$d$ electron the binding energy of the Ti 2$p$ core level of Ti$^{3+}$-ions is reduced with respect to that of Ti$^{4+}$-ions, making this core level an excellent measure of the 3$d$ band-filling. The measurements were carried out in situ, i.e., keeping the samples in ultra high vacuum after PLD growth, to avoid any post-fabrication oxidation. As is clear from Figure \ref{Fig1}(d), all samples suffer from strong over-oxidation. Even for fabrication in vacuum ($p<2\times10^{-8}$ mbar) the dominating valency is Ti$^{4+}$, corresponding to an electronic occupancy far away from half band-filling.
Similar to the La$_2$Ti$_2$O$_7$ phase, breaking the shared corners of adjacent TiO$_6$ octahedra allows the incorporation of excess oxygen ions but---in contrast to La$_2$Ti$_2$O$_7$---only locally and in a random manner as sketched for the LaTiO$_{3+x}$ case in Figure \ref{Fig1}(a). Such randomly distributed excess oxygen does not result in coherent Bragg peaks, making diffraction techniques insensitive to this kind of over-oxidation. Therefore a direct characterization of the electronic properties is crucial and an analysis of the merely structural properties may be misleading.

At this point, we briefly comment on the implications of these results for oxide heterostructures, in which LTO is frequently integrated as an electron donor layer.$^\text{\cite{disa_orbital_2015,cao_engineered_2016}}$ In such structures special care about the titanium valency should be exercised, since the observation of Ti$^{4+}$ in the LTO layer may not prove the intended electron transfer. It rather could signal the over-oxidation discussed above.

Avoiding over-oxidation, i.e., gaining control of the oxygen stoichiometry in LTO thin films is an essential prerequisite for controlled $p$-doping. Due to the elevated temperatures required during thin film fabrication substantial oxygen diffusion between substrate and film is expected.$^\text{\cite{schneider_origin_2010}}$ Since STO is prone to oxygen vacancy formation upon vacuum annealing and exhibits a high oxygen diffusion coefficient, we expect a strong oxygen out-diffusion from this substrate material.$^\text{\cite{dudy_situ_2016,schneider_origin_2010,paladino_oxygen_1965,henrich_surface_1978,posadas_scavenging_2017}}$ This hypothesis is tested by introducing an epitaxial LaAlO$_{3-x}$ (LAO) buffer layer between substrate and LTO film to block the oxygen out-diffusion.
As depicted in Figure \ref{Fig2}(a), these samples indeed exhibit an increased Ti$^{3+}$ signal (black curve) but still show a dominant Ti$^{4+}$ contribution. The latter can be further reduced by employing DyScO$_3$ (DSO) substrates such that the Ti$^{3+}$ signal is prevailing.
Upon reducing the Ti$^{4+}$ content, the structural properties as characterized by RHEED and LEED further improve. A layer-by-layer growth mode and an atomically flat and clean LTO surface is achieved for fabrication on DSO substrates as discussed in detail in Section S1 (Supporting Information). Note that for epitaxy on DSO substrates the best structural properties and lowest Ti$^{4+}$ content are achieved for growth in an oxygen background pressure of $5\times10^{-8}\;$mbar, and not in vacuum as for STO based samples (see Figure S2, Supporting Information). Without a sizable oxygen out-diffusion from the DSO substrate, additional oxygen needs to be supplied to reach the optimal oxidation conditions for the growth of LTO thin films.

The origin of the remaining over-oxidation on DSO based samples is of a different nature, as revealed by angle-dependent XPS measurements presented in Figure \ref{Fig2}(c).
Increasing the photoelectron emission angle $\vartheta$ towards grazing emission---and thereby decreasing the effective probing depth
$\lambda_{eff} = \lambda_{0}~\text{cos}~\vartheta$---results in a strong enhancement of the Ti$^{4+}$ signal.
This demonstrates that the remaining over-oxidation is limited to a strongly confined surface layer on the length scale of the photoelectron inelastic mean free path $\lambda_0=18.9\;\text{\AA}$.$^\text{\cite{sing_profiling_2009}}$
A plausible origin for this surface over-oxidation is the adsorption of additional oxygen ions at the apical sites completing the surface TiO$_6$ octahedra and removing two electrons per anion from the LTO film, as sketched in Figure \ref{Fig2}(d). Such apical oxygen ligands have already been detected by scanning tunneling microscopy on the surface of the related perovskite SrVO$_3$.$^\text{\cite{okada_quasiparticle_2017}}$
A quantitative evaluation of the Ti$^{3+}$ to Ti$^{4+}$ intensity ratio supports this scenario and is plotted in Figure \ref{Fig2}(e). Due to an apical oxygen coverage $\alpha$, a surface layer containing Ti$^{4+}$-ions with an extension of $d$ unit cells is assumed. The angle-dependent XPS data is best reproduced for $d=2$ uc and a coverage $\alpha \approx$ 75\%. Details of the model calculation can be found in Section S3 (Supporting Information).
\begin{figure*}[bpht]
{
\includegraphics[width = 1\textwidth]{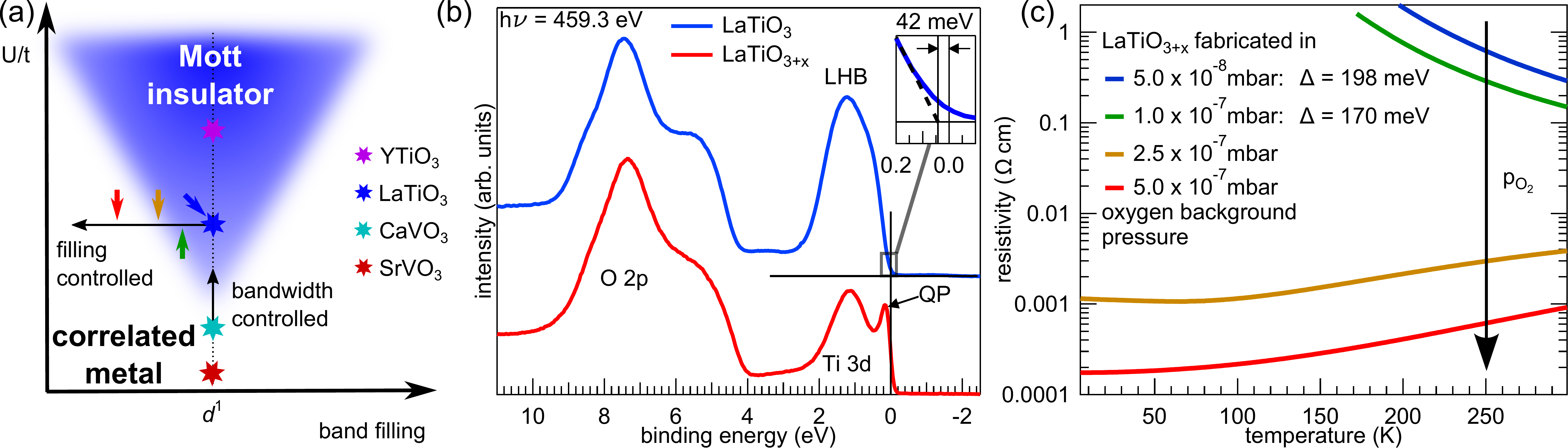}
}
\caption{
\label{Fig3}
(a) Sketch of the electronic phase diagram of a Mott system.
(b) Resonant photoemission spectra of stoichiometric LaTiO$_3$ and $p$-doped LaTiO$_{3+x}$ thin films measured at the Ti $L$ absorption edge.
(c) Temperature dependent resistivity measurements of LTO thin films fabricated at various oxygen growth pressures. Photoemission and transport measurements demonstrate the transition from Mott insulator to correlated metal as the oxygen growth pressure is increased, i.e., upon $p$-doping. The corresponding tentative positions in the electronic phase diagram are marked in (a) by arrows in a consistent color code.
}
\end{figure*}

These apical oxygen sites can be blocked by an epitaxial LAO capping layer.
Since the cations in the LAO capping layer are of the same oxidation states as the cations in bulk LTO, the situation for the topmost TiO$_2$ layer (the former surface layer) is similar to bulk LTO. The oxygen octahedra are completed and the lanthanum and aluminum cations of the capping layer supply their share of electrons, thereby avoiding an electron depletion from Ti$^{3+}$-ions (see Figure \ref{Fig2}(d)).
LAO-capped LTO thin films grown on DSO substrates indeed exhibit no Ti$^{4+}$ signal indicating a stoichiometric LaTi$^{3+}$O$_3$ thin film (see green spectrum in Figure 2(c)). Furthermore the capping layer prevents any further oxidation upon exposure to air which is crucial for future device fabrication (see Section S4, Supporting Information).

We like to point out here that the complex over-oxidation mechanisms in LTO thin films have recently led to confusion about the line shape of the 2$p$ spectrum in $d^1$ transition metal oxides. It has been argued that photoemission final states always lead to a Ti$^{4+}$ signal even for stoichiometric LTO exclusively comprised of Ti$^{3+}$-ions.$^\text{\cite{lin_final-state_2015}}$ Our measurements, however, give proof that this is not the case. Rather, a pure Ti$^{3+}$ spectrum without any detectable additional satellite structures is observed. For reference the Ti 2$p$ photoemission and the Ti $L$ edge x-ray absorption spectra of stoichiometric LTO thin films are provided in Figure S5 (Supporting Information).

Having demonstrated how to gain control of the oxygen stoichiometry, we now deliberately tune the over-oxidation to drive the filling controlled MT.

The electronic phase diagram as function of correlation strength $U/t$ vs. band-filling is schematically depicted in Figure \ref{Fig3}(a), where $U$ is the on-site Coulomb repulsion and $t$ the intersite hopping integral. For orientation also the series of prototypical $d^1$-materials discussed above is included. The different members (colored stars) are situated on different positions along the band\textit{width} controlled MT, which is depicted by the vertical arrow.
Reducing the band-\textit{filling} by supplying excess oxygen allows us to continuously change the position along the filling controlled MT, marked by the horizontal arrow and starting from stoichiometric LTO (blue star) in Figure \ref{Fig3}(a). As shown above, additional oxygen can easily be incorporated in LTO thin films by increasing the oxygen pressure applied during thin film fabrication. Every excess oxygen ion removes two electrons from the titanium 3$d$ band, which eventually leads to a transition into the correlated metal phase.

We employ resonant photoemission at the Ti $L$ absorption edge to monitor the Ti 3$d$ spectral function across the transition. Note that the samples were measured in situ and without an LAO capping. The photoelectron inelastic mean free path in this experiment is high enough to probe the bulk LTO film below the over-oxidized surface layer. For the employed photon energy of $460\;$eV, $\lambda_0=10.6\;\text{\AA}$ and the relative contribution of the bulk signal is about 50$\%$. Furthermore, the resonant conditions strongly and selectively enhance the Ti 3$d$ photoemission signal. The resulting spectra of a stoichiometric and a $p$-doped LTO thin film are depicted in Figure \ref{Fig3}(b). In the stoichiometric case a single Ti 3$d$ feature is observed near the chemical potential and identified as the lower Hubbard band (LHB) characteristic for the Mott insulating phase. Interestingly, the over-oxidized surface layer on these bare LTO samples does not lead to metallic charge carriers, signalled by the absence of a Fermi cutoff. The most plausible scenario is that the surface layer rather forms areas with a band-filling of $d^0$, i.e., with no contribution to the Ti 3$d$ spectrum, and subjacent or adjacent stoichiometric areas in the Mott insulating phase with a $d^1$ band-filling. Thus, no trace of intermediately doped (i.e. metallic) areas is detected. The onset of the LHB is found at a binding energy of $42\;$meV, which is a lower limit for the Mott gap between the upper and the lower Hubbard band. The presence of a lower Hubbard band and a gap size of at least $42\;$meV indicate that the Ti 3$d$ electrons are fully localized.

For the $p$-doped sample, labeled LaTiO$_{3+x}$ in Figure \ref{Fig3}(b), a Fermi cutoff and the typical two peak structure of a correlated metal, consisting of a coherent quasi-particle peak (QP) and an incoherent LHB, is observed.
These findings demonstrate that fabrication of LTO thin films at elevated oxygen pressure is a suitable method to reduce the band-filling and trigger the MT.

To directly probe the transport properties, which are crucial for device applications, we have also performed complementary temperature dependent resistivity measurements. As depicted in Figure \ref{Fig3}(c), an activated behavior is observed for stoichiometric samples fabricated in an oxygen pressure of $5\times10^{-8}\;$mbar. An Arrhenius-type temperature dependence of the resistivity with $\rho \propto \exp{\frac{\Delta}{2k_BT}}$ reproduces the data and yields a Mott gap of $\Delta = 198\;$meV. Upon $p$-doping by means of increasing the oxygen growth pressure, the gap initially decreases to $\Delta = 170\;$meV for growth in $1\times10^{-7}\;$mbar of oxygen before it collapses for higher oxygen pressures as indicated by the metal-like temperature dependence. The change of resistivity is drastic as the phase boundary is crossed, being a factor of approx. 100 at room temperature and even several orders of magnitude higher at low temperatures. The resulting positions of these samples in the electronic phase diagram are marked by arrows in the corresponding colors in Figure \ref{Fig3}(a).

The characteristics of the MT described above make the material an exciting candidate for Mott transistors. Samples on the insulating side of the phase transition exhibit a Mott gap of about $200\;$meV, large enough to avoid thermally activated transport crucial for a well defined OFF-state. The phase transition is easily reached by $p$-doping and yields a change of orders of magnitude in conductivity. In addition, the material's position relative to the band-filling induced MT can be adjusted by an easily controlled fabrication parameter, i.e., the oxygen pressure.

Besides the band-filling control demonstrated above, fabrication of LTO as thin films allows for further tuning of the electronic properties. For instance, the bandwidth can be effectively reduced by decreasing the film thickness in favor of the Mott insulating phase.$^\text{\cite{yoshimatsu_dimensional-crossover-driven_2010}}$ Strain engineering and thereby manipulating the in-plane lattice parameter is another approach and works both ways---increasing and decreasing $t$ allows
 to shift the material towards the correlated metal and Mott insulating phase, respectively.$^\text{\cite{yoshimatsu_strain-induced_2016}}$

The possibility to relocate the material's position in the electronic phase diagram is especially important for the optimization of the operating point of future Mott transistor devices. As discussed above, approaching the phase transition from the insulating side is predicted to make the material susceptible to moderate electric fields, thereby allowing for conventional field effect setups.

Turning from the high controllability of the electronic properties in LTO thin films to the practical realization of a field effect based Mott transistor, the material's perovskite crystal structure is advantageous. This material class offers many epitaxially compatible functional compounds, ready for the design of all-oxide devices. For instance the two main components of a field effect device, dielectric and electrode, can be realized by the high-$\kappa$ material STO and highly conductive SrRuO$_3$ or SrVO$_3$, respectively.$^\text{\cite{okhay_manipulation_2013,moyer_highly_2013}}$

In summary, we have established a suitable method to control the band-filling in the prototypical Mott insulator LTO by excess oxygen doping. The films can readily accommodate excess oxygen ions, which was demonstrated as an effective way of $p$-doping to control the material's electronic properties. The choice of the substrate was found to be crucial to gain control of the oxygen stoichiometry. Films grown on STO are inherently over-oxidized, while fully stoichiometric LTO thin films were fabricated on DSO substrates. These are important steps in material development towards the realization of a Mott transistor, which elegantly utilizes the concept of conventional field effect setups, while harnessing at the same time the peculiar electronic properties caused by electron correlations.



\section*{Experimental Section}
\textit{Substrate termination:} A TiO$_2$ surface termination of STO (001) substrates was achieved by rinsing in deionized water, etching with
hydrofluoric acid, and subsequent annealing in oxygen as described elsewhere.$^\text{\cite{koster_quasi-ideal_1998}}$  ScO$_2$-terminated surfaces of DSO (110) (orthorhomic notation) were prepared by annealing in oxygen, followed by etching in NaOH.$^\text{\cite{kleibeuker_structure_2012}}$ All substrates were supplied by CrysTec GmbH.\\

\textit{LaTiO$_3$ thin film epitaxy by pulsed laser deposition (PLD):}
A PLD system with a base pressure in the 10$^{-10}\;$mbar range was used to fabricate the LTO thin films. To stabilize oxygen pressures in the 10$^{-8}\;$mbar range at elevated temperatures, a UHV anneal of the substrates at 500$\;^{\circ}$C was performed prior to the ablation process. The substrate temperature during ablation was held at $T_S=800\;^{\circ}$C. Temperature control was realized by an infrared laser ($\lambda = 980\;$nm) and a two color pyrometer, both directed to the back side of the sample holder.
The LTO material was ablated from a polycrystalline LaTiO$_{3+x}$ target by a KrF excimer laser ($\lambda = 248\;$nm) with a laser fluency of $\Phi=1.5\;$J/cm$^2$ at a repetition rate of $f=1\;$Hz and a target-to-substrate distance of $54\;$mm.
The resulting growth rate was 30 pulses per unit cell, as monitored by RHEED intensity oscillations. The thickness of the fabricated LTO thin films was between 20 and 40 uc.
For samples fabricated in oxygen background pressure, the oxygen dosing valve was closed directly after the ablation process and the samples were cooled down to room temperature in vacuum. The LAO capping layers were ablated from a single crystalline LaAlO$_3$ target with $\Phi=1.3\;$J/cm$^2$ and $f=1\;$Hz in an oxygen background atmosphere of $5\times10^{-8}\;$mbar. During epitaxy of the capping the substrate was held at $T_S=700\;^{\circ}$C.

\textit{LTO thin film characterization:}
The PLD chamber is attached to a UHV cluster with a base pressure in the $10^{-11}\;$mbar range, allowing for an in situ characterization by XPS and LEED directly after the PLD growth.
The XPS setup utilizes a monochromatized Al K$_{\alpha}$ x-ray source and an Omicron EA-125 electron analyzer with an overall energy resolution of approx. 0.40$\;$eV.
Resonant photoemission measurements were performed at the two-color beamline I09 at DIAMOND Light source using linearly polarized light with a photon energy of 459.3$\;$eV. Hard x-ray photoemission was performed on the same spot (size 30$\;\mu$m x 50$\;\mu$m) with a photon energy of 3$\;$keV. The spectra were recorded with an EW4000 photoelectron analyzer (VG Scienta, Uppsala, Sweden) equipped with a wide-angle acceptance lens. A portable UHV chamber with a base pressure in the 10$^{-9}\;$mbar range equipped with a NEG pump was used to ship the samples to DIAMOND light source and prevent oxidation during transport. Temperature dependent resistivity measurements were performed in the Van-der-Pauw geometry using a Quantum Design Physical Properties Measurement System (PPMS) on LTO thin films capped with 5 uc LAO to prevent oxidation in air. Electrical contacts on the LTO thin films were realized by ultrasonic wire bonding with Al wires.

\section*{Supporting Infromation}

Supporting Information is available online or from the authors.

\section*{Acknowledgements}

The authors are grateful for funding support from the
Deutsche Forschungsgemeinschaft (FOR 1162 and SFB 1170 "ToCoTronics")
and acknowledge Diamond Light Source for time on beamline
I09 under proposal SI14106, SI15200, SI15856, and NT18372. The authors would also like to thank D. McCue for
his superb technical support at the I09 beamline. We also acknowledge fruitful discussion with M. Bibes and R. Aeschlimann.

\bibliographystyle{apsrev4-1}

\end{document}


\preprint{APS/123-QED}

\title{Supporting Information\\ Tailoring Materials for Mottronics: Excess Oxygen Doping of a Prototypical Mott Insulator}

\author{P. Scheiderer}
\affiliation{Physikalisches Institut and R\"ontgen Center for Complex Material Systems (RCCM), Universit\"at W\"urzburg, Am Hubland,
D-97074 W\"urzburg, Germany}
\author{M. Schmitt}
\affiliation{Physikalisches Institut and R\"ontgen Center for Complex Material Systems (RCCM), Universit\"at W\"urzburg, Am Hubland,
D-97074 W\"urzburg, Germany}
\author{J. Gabel}
\affiliation{Physikalisches Institut and R\"ontgen Center for Complex Material Systems (RCCM), Universit\"at W\"urzburg, Am Hubland,
D-97074 W\"urzburg, Germany}
\author{M. Zapf}
\affiliation{Physikalisches Institut and R\"ontgen Center for Complex Material Systems (RCCM), Universit\"at W\"urzburg, Am Hubland,
D-97074 W\"urzburg, Germany}
\author{M. St\"ubinger}
\affiliation{Physikalisches Institut and R\"ontgen Center for Complex Material Systems (RCCM), Universit\"at W\"urzburg, Am Hubland,
D-97074 W\"urzburg, Germany}
\author{\\P. Sch\"utz}
\affiliation{Physikalisches Institut and R\"ontgen Center for Complex Material Systems (RCCM), Universit\"at W\"urzburg, Am Hubland,
D-97074 W\"urzburg, Germany}
\author{L. Dudy}
\affiliation{Physikalisches Institut and R\"ontgen Center for Complex Material Systems (RCCM), Universit\"at W\"urzburg, Am Hubland,
D-97074 W\"urzburg, Germany}
\author{C. Schlueter}
\affiliation{Diamond Light Source Ltd., Didcot, Oxfordshire OX11 0DE, United Kingdom}
\author{T.-L. Lee}
\affiliation{Diamond Light Source Ltd., Didcot, Oxfordshire OX11 0DE, United Kingdom}
\author{M. Sing}
\affiliation{Physikalisches Institut and R\"ontgen Center for Complex Material Systems (RCCM), Universit\"at W\"urzburg, Am Hubland,
D-97074 W\"urzburg, Germany}
\author{R. Claessen}
\affiliation{Physikalisches Institut and R\"ontgen Center for Complex Material Systems (RCCM), Universit\"at W\"urzburg, Am Hubland,
D-97074 W\"urzburg, Germany}

\date{\today}

\maketitle

\subsection{Structural characterization}
{
In the main text, the impact of the substrate choice on the oxygen content in LTO thin films was probed by the titanium valency as determined from the Ti $2p$ core level photoemission spectrum. The strong oxygen-out diffusion from STO substrates causes an over-oxidation of the LTO thin films. This diffusion is partially blocked by an LAO buffer layer and absent when DSO substrates are used. These findings can be correlated with information from a comprehensive structural characterization.

As depicted in Figure \ref{FigS1}(a), the RHEED intensity oscillations of the specular reflex are strongly damped for epitaxy on STO substrates. Such a behavior indicates a roughening of the LTO thin film surface as the film growth process progresses. This scenario is supported by the RHEED and LEED diffraction patterns of the LTO surface after the PLD process, displayed in Figure \ref{FigS1}(b), red box. The diffraction patterns are blurry and the Bragg reflexes in LEED are connected by faint streaks (highlighted by the dashed lines), typical for a defect rich surface. The situation changes when an LAO buffer layer is introduced between the LTO thin film and the STO substrate. The buffer layer is fabricated under the same conditions as the LTO thin film (residual gas pressure $<2\times10^{-8}\;$mbar and $T_S$=800$^{\circ}$C). The subsequent fabrication of the LTO thin film displays regular RHEED intensity oscillations and the quality of the RHEED and LEED diffraction patterns is significantly improved. Further improvement of the diffraction pattern quality is obtained for growth on DSO substrates. The enhanced structural quality correlates with the changes of the titanium valency towards the correct Ti$^{3+}$ oxidation state, as presented in Figure 2 of the main text. The LTO thin films fabricated on DSO substrates have a high structural quality as judged by the RHEED and LEED patterns and are atomically flat as determined by atomic force microscopy measurements (see Figure \ref{FigS1} (c)).
\begin{figure*}[hbpt]
{
\includegraphics[width = 0.9\textwidth]{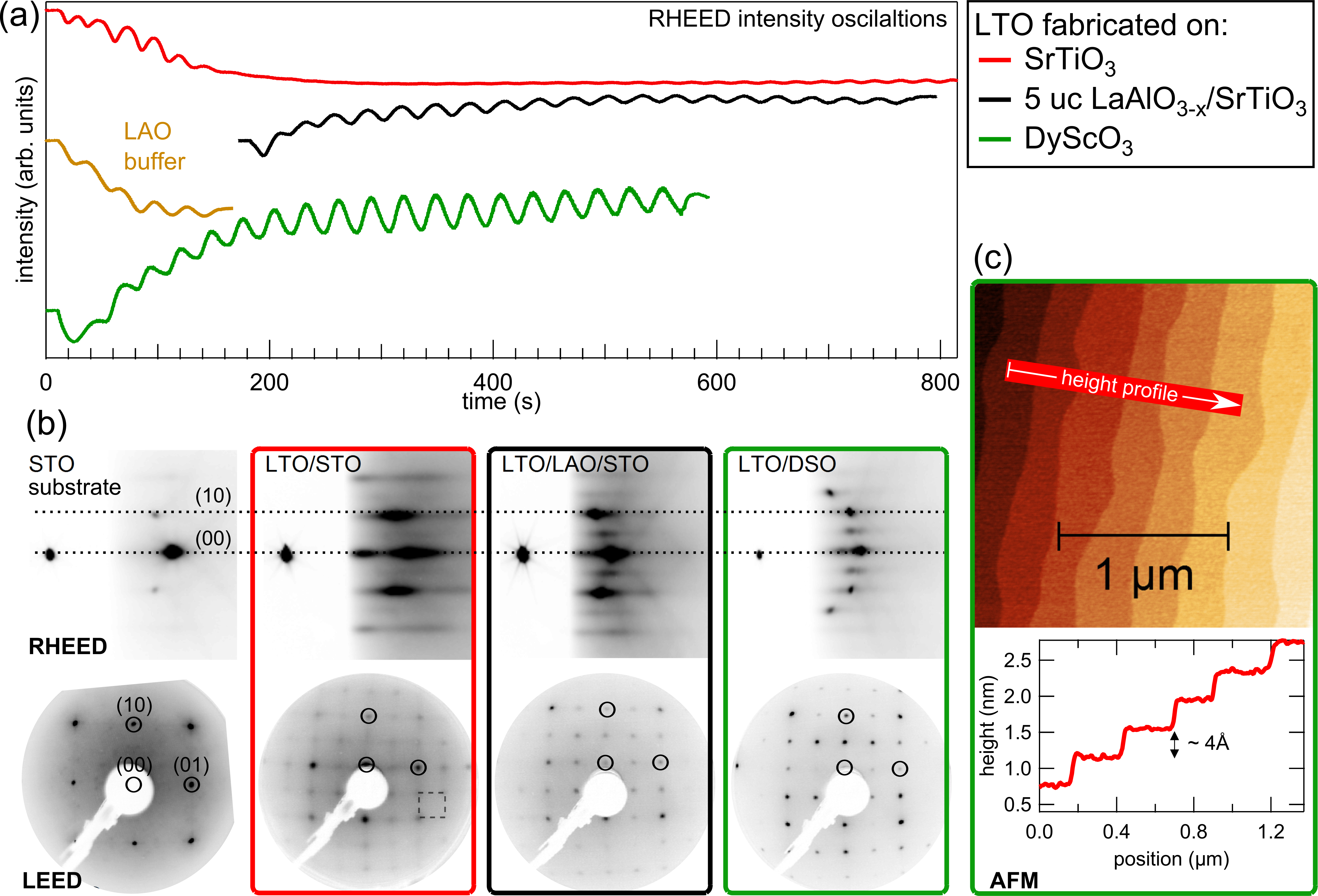}
}
\caption{
\label{FigS1}
Structural characterization for thin LTO films fabricated on STO, LAO buffered STO and DSO substrates. The color code is consistent for all sub-figures. (a) RHEED intensity oscillations of the specular reflex for LTO growth on different substrates. Epitaxy on STO substrates results in strongly damped oscillations, whereas on LAO buffered STO and on DSO substrates regular oscillations are observed, indicating a layer-by-layer growth mode. (b) RHEED and LEED diffraction patterns of the corresponding LTO surfaces together with diffraction patterns from the STO surface for comparison. The quality of the patterns obtained improves as the correct titanium valency is approached. The dashed lines in the red box highlight the streaks between the Bragg peaks. The dotted lines are a guide to the eye. (c) AFM measurements on an LTO film grown on a DSO substrate indicate an atomically flat surface with a step height of a single unit cell.
}
\end{figure*}
}
\newpage

\subsection{Optimization of oxygen growth pressure for DyScO$_3$ based samples}
{
For growth of LTO thin films on STO substrates it was shown in the main text that the best structural properties and the highest amount of Ti$^{3+}$ is obtained for fabrication at the lowest achievable oxygen pressures, i.e., base pressure conditions of the chamber ("in vacuum"). Therefore the optimal oxidation state is not accessible even for the lowest achievable oxygen growth pressures in our setup.

The situation is different, when DSO substrates are used instead of STO. As depicted in Figure \ref{FigS2} (a), the highest Ti$^{3+}$ to Ti$^{4+}$ ratio is obtained for an oxygen growth pressure of $5\times10^{-8}\;$mbar. For fabrication in vacuum (residual gas pressure $p<2\times10^{-8}\;$mbar) even a Ti$^{2+}$ signal is detected and the Ti$^{3+}$ to Ti$^{4+}$ intensity ratio is strongly decreased. Higher oxygen pressures in the $10^{-7}\;$mbar range lead to a decrease of the Ti$^{3+}$ to Ti$^{4+}$ ratio.

The corresponding LEED patterns depicted in Figure \ref{FigS2} (b-e) indicate the structural quality of the LTO thin film surfaces.
Again, the highest structural quality is obtained for an oxygen growth pressure of $5\times10^{-8}\;$mbar, as judged from the contrast and width of the Bragg spots.
Fabrication in vacuum yields only a faint LEED pattern (Figure \ref{FigS2} (b)). For higher oxygen growth pressures $p_{O_{2}}>5\times10^{-8}\;$mbar the LEED pattern is still well pronounced, but a gradual decrease of the structural quality can be recognized as $p_{O_{2}}$ is increased. Based on these findings, we identify p$_{O_{2}}=5\times10^{-8}\;$mbar as the optimal value for the fabrication of LTO thin films on DSO substrates.

\begin{figure*}[hbpt]
{
\includegraphics[width = 1\textwidth]{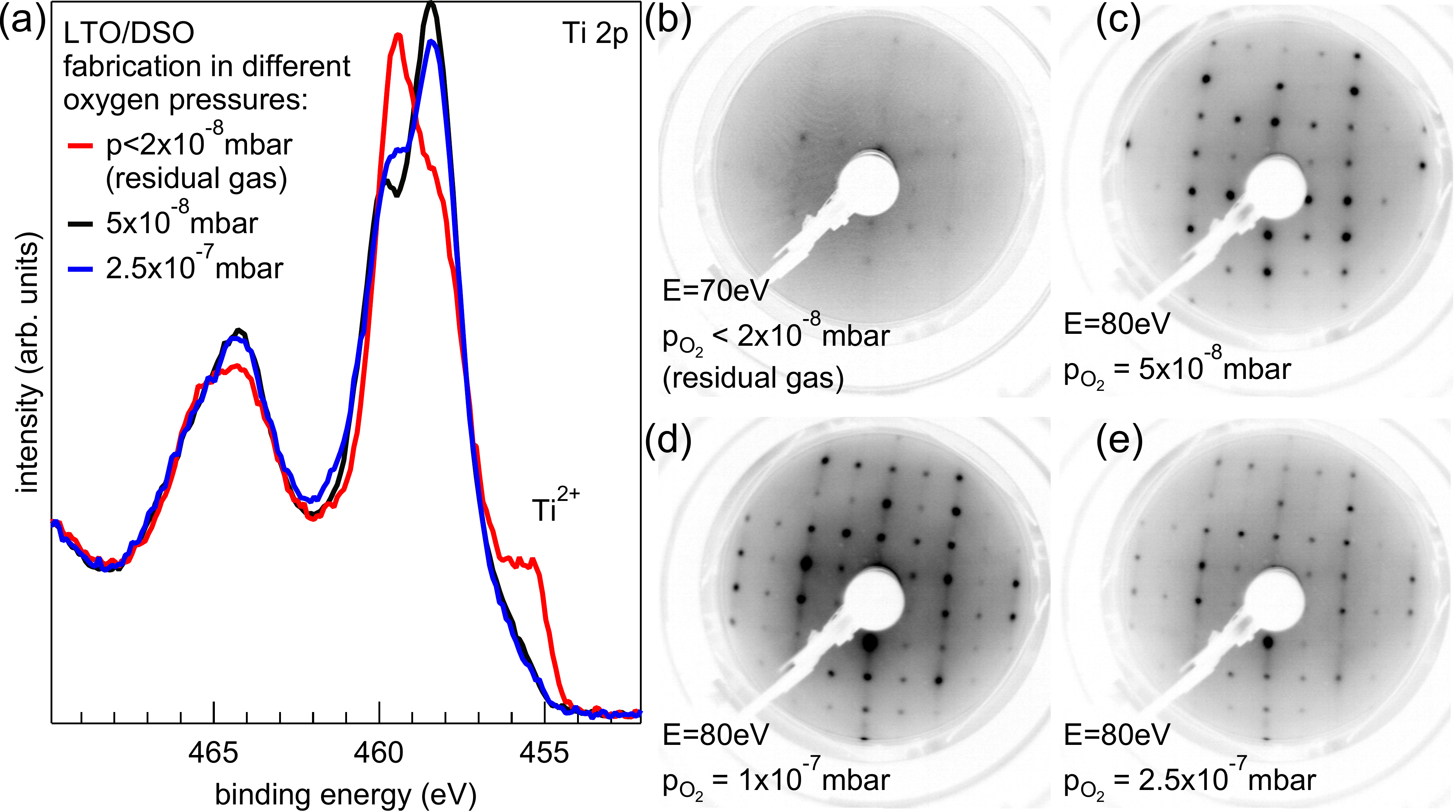}
}
\caption{
\label{FigS2}
Optimization of oxygen growth pressure for LTO thin film fabrication on DSO substrates. (a) In situ XPS measurements of the Ti $2p$ core level. The highest Ti$^{3+}$ to Ti$^{4+}$ intensity ratio is obtained for p$_{O_2}=5\times10^{-8}\;$mbar. For lower oxygen pressures Ti$^{2+}$ is detected and an increased Ti$^{4+}$ content is observed. Higher oxygen pressures induce a gradual decrease of the Ti$^{3+}$ to Ti$^{4+}$ ratio. (b-c) LEED patterns of LTO thin films fabricated in the indicated oxygen pressure. The quality of the LEED patterns correlates with the Ti$^{3+}$ to Ti$^{4+}$ intensity ratio and indicates an optimal oxygen growth pressure of $5\times10^{-8}\;$mbar for LTO thin film fabrication on DSO substrates.
}
\end{figure*}
}
\newpage

\subsection{Surface over-oxidation of LaTiO$_3$/DyScO$_3$}
{
The angle dependent XPS measurements of the Ti $2p$ core level of LTO/DSO were fitted by a superposition of Ti$^{4+}$ and Ti$^{3+}$ reference spectra with variable weight factors for each reference spectrum. The Ti$^{4+}$ reference spectrum was obtained from a strongly over-oxidized LTO thin film fabricated on STO, which shows a negligible Ti$^{3+}$ signal. The Ti$^{3+}$ reference was measured on an LAO-capped stoichiometric LTO/DSO sample. As depicted in Figure \ref{FigS3}(a-d), the experimental data is well reproduced by the fits. The resulting Ti$^{3+}$ to Ti$^{4+}$ ratios are plotted in Figure \ref{FigS3}(e) and are compared to a microscopic model. As sketched in Figure \ref{FigS3}(f), the model assumes a Ti$^{4+}$-containing surface layer with an extension of $d$ unit cells. The over-oxidation is assumed to be caused by adsorbed apical oxygen with a relative surface coverage $\alpha$. Every apical oxygen removes two electrons from the Ti $3d$ subshell and thereby generates two Ti$^{4+}$ ions. We assume a constant relative Ti$^{4+}$ content $x$ in the surface layer, expressed by $x=\frac{2\alpha}{d}$. Based on this model and considering the exponential damping caused by inelastic scattering of the photoelectrons, the photoemission intensity ratio between the Ti$^{3+}$ and Ti$^{4+}$ signal is given by:
\begin{equation}
\frac{I(Ti^{3+})}{I(Ti^{4+})}=\frac{\sum^{d-1}_{j=0}~(1-\frac{2\alpha}{d})~e^{-\frac{j \cdot c}{\lambda_0 cos{\vartheta}}}+\sum^{\infty}_{j=d}~e^{-\frac{j \cdot c}{\lambda_0 cos{\vartheta}}}}{\sum^{d-1}_{j=0}~\frac{2\alpha}{d}~e^{-\frac{j \cdot c}{\lambda_0 cos{\vartheta}}}}.
\end{equation}
The out-of-plane lattice constant was set to $c=4\;\text{\AA}$ and the photoelectron inelastic mean free path $\lambda_0$ was determined by the empirical formula of Tanuma, Penn, and Powell (TPP-2M) and amounts to $18.9\text{ \AA}$.\cite{tanuma_calculations_1994} The fits to the measured data were performed by converging to the best values of $\alpha$ at several fixed values of $d$, the results are displayed in Figure \ref{FigS3}. For $d=1$ the model reaches the limit of a pure Ti$^{4+}$ surface layer at a coverage of $\alpha=0.5$ and clearly overestimates the Ti$^{3+}$ to Ti$^{4+}$ intensity ratio. This demonstrates, that the extension of the surface layer exceeds one unit cell. The best fit is obtained for $d=2$ at a coverage of $\alpha=0.7$ and reproduces the absolute value of the Ti$^{3+}$ to Ti$^{4+}$ intensity ratio as well as its angular dependence. Higher values for $d$ underestimate the variation of the intensity ratio with the emission angle $\vartheta$ and the fits exhibit a larger $ \mathcal{X}^2$ value. This analysis provides strong support that the remaining over-oxidation of LTO/DSO samples is strongly confined to the sample surface, extending only two unit cells into the sample. Due to the quantitative agreement between the microscopic model and the measured angle dependent XPS data we identify adsorbed apical oxygen ions as the reason for the remaining over-oxidation. The exact coverage $\alpha$ may delicately depend on the base pressure in the vacuum chamber and the storage time, i.e. the oxygen dose the sample is exposed to between fabrication and measurement. More oxygen can be adsorbed from the residual gas, probably until a coverage of $\alpha=1$. An effective way to avoid or remove this kind of over-oxidation is an LaAlO$_{3-x}$ capping layer as demonstrated in the main text.
\begin{figure*}[hbpt]
{
\includegraphics[width = 1\textwidth]{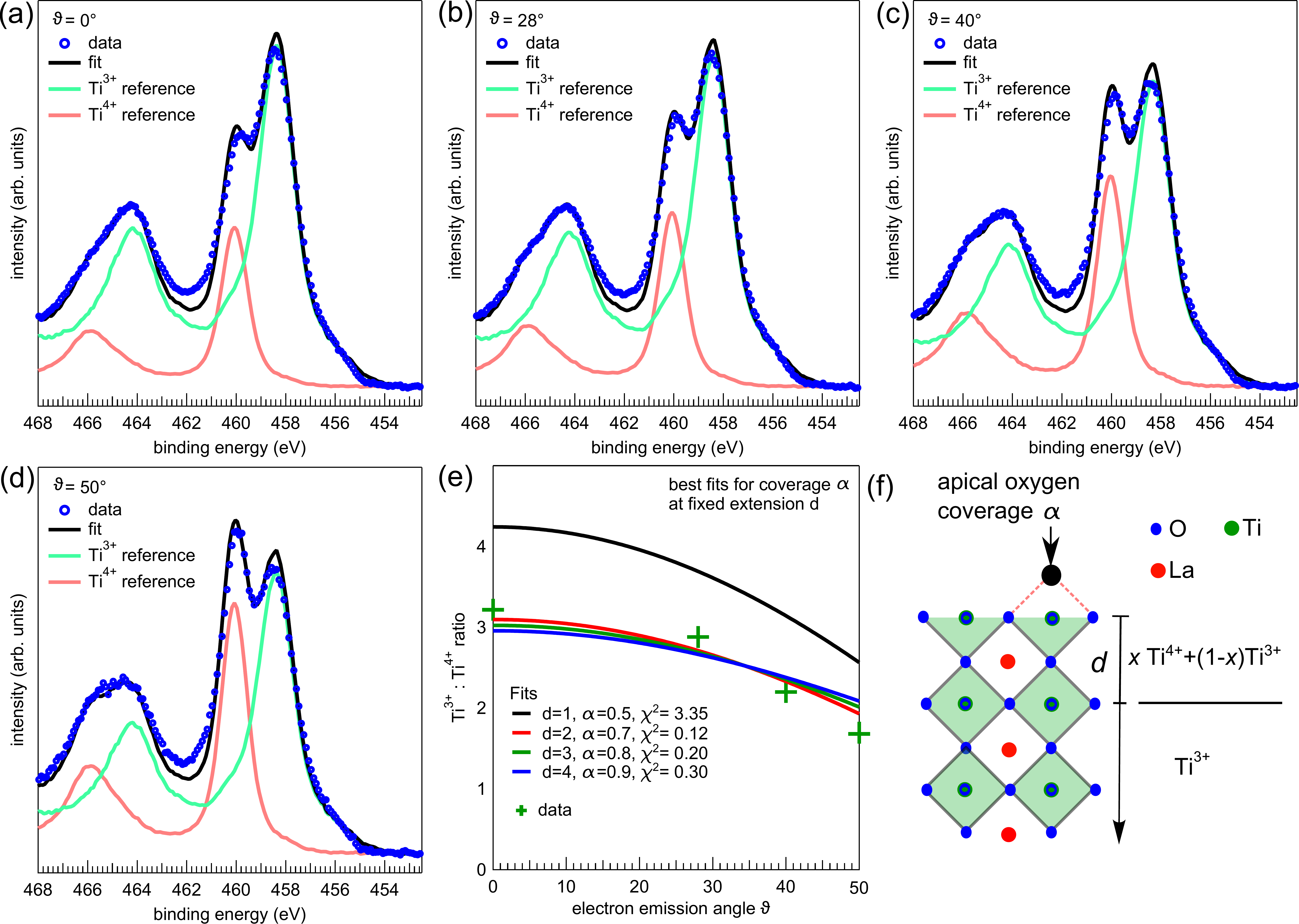}
}
\caption{
\label{FigS3}
Angle-dependent XPS spectra of the Ti $2p$ core level at various electron emission angles $\vartheta$ measured on a bare LTO thin film fabricated on DSO. (a-d) The spectra are fitted by a superposition of Ti$^{3+}$ and Ti$^{4+}$ reference spectra.
(e) The resulting Ti$^{3+}$ to Ti$^{4+}$ ratio is used to evaluate the coverage $\alpha$ by apical oxygen and the extension $d$ of the over-oxidized surface layer within a simple model (see text for details). The data is best reproduced for a coverage of approx. 75\% and an extension of $d = 2\;$uc.
(f) Simplified sketch of the microscopic situation expected to cause the over-oxidation at the LTO surface.
}
\end{figure*}
}

\newpage
$~$
\newpage
%

\subsection{Passivation of the LaTiO$_3$ thin film surface}
{

As discussed in the main text, LTO is prone to over-oxidation. In view of transport measurements or future device fabrication, sample handling in air is inevitable.
Therefore, we checked for changes of the titanium valency when samples are exposed to air. To this end a Ti $2p$ spectrum was measured on a stoichiometric LTO thin film in situ, i.e. keeping the sample in ultrahigh vacuum. The sample was then removed from the vacuum system and stored in air for several days and then put back into the vacuum system to repeat the XPS measurement.The results are depicted in Figure \ref{FigS5}(a). The titanium valency has clearly changed and only a faint Ti$^{3+}$ signal remains after storage in air (ex situ). Thus a method to protect the LTO thin films from over-oxidation in air is required.

In Figure 2 of the main text, it was demonstrated that an epitaxial LAO capping layer blocks the adsorption sites for apical oxygen that occurs on the bare LTO surface. Such an LAO capping layer can also be employed to protect LTO thin films from over-oxidation. To demonstrate the successful passivation, XPS measurements on stoichiometric LTO thin films with a 5 unit cell thick LAO capping layer were again performed in situ and ex situ. The corresponding Ti $2p$ spectra are displayed in Figure \ref{FigS5} (b). No changes in the Ti $2p$ spectrum are observed, indicating that the titanium valency remains unchanged upon exposure to air.

\begin{figure*}[hbpt]
{
\includegraphics[width = 1\textwidth]{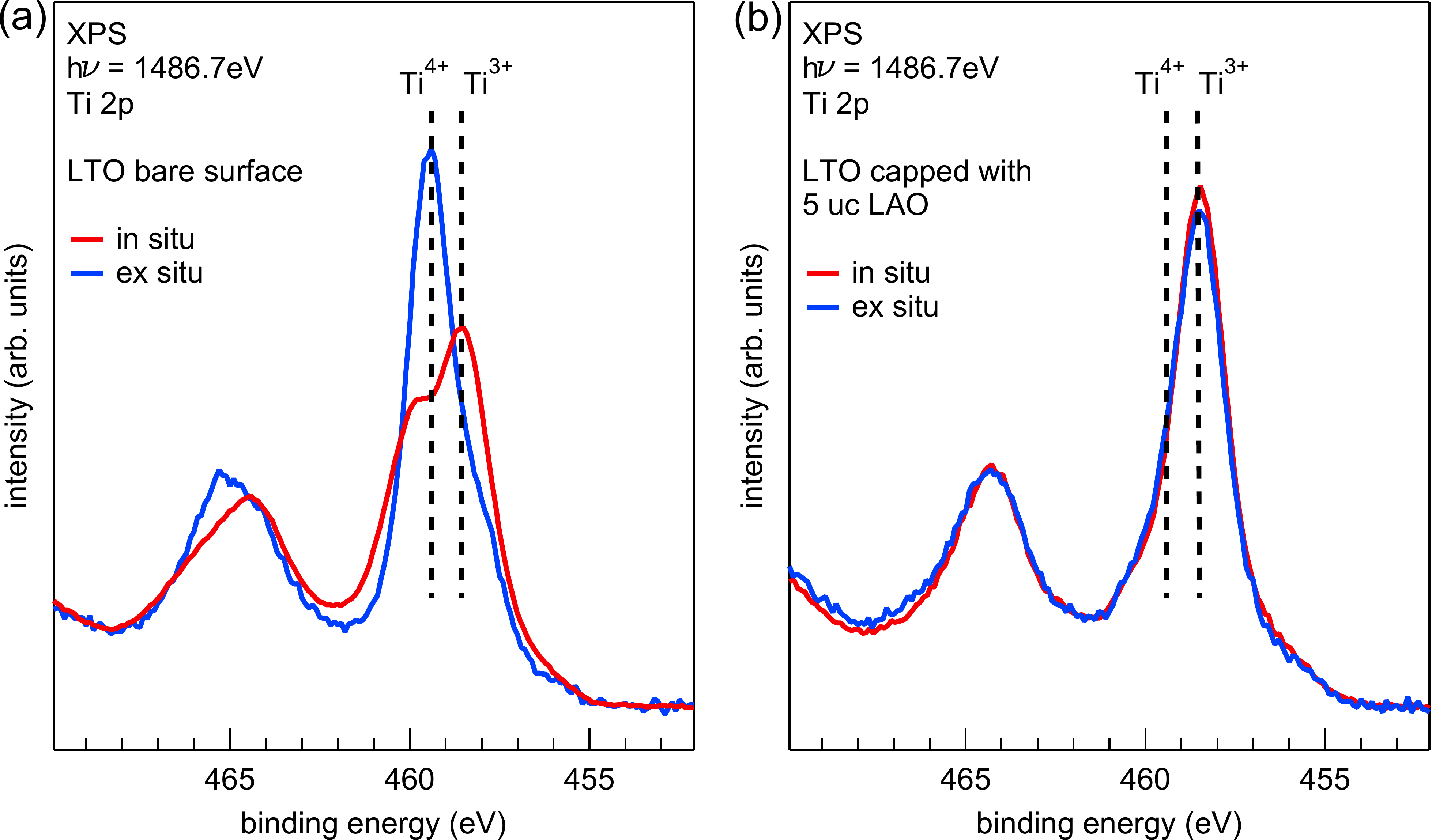}
}
\caption{
\label{FigS5}
XPS measurements of the Ti 2$p$ core level before (in situ) and after (ex situ) exposure to air performed on bare LTO films (a) and LAO-capped LTO films (b).
(a) Without a protecting capping layer, the LTO thin films strongly over-oxidize during storage in air. (b) A 5 uc thick LAO capping layer fully passivates the LTO thin films, no changes in the Ti $2p$ spectrum are observed between the in situ and ex situ measurement.
}
\end{figure*}
}
\newpage
\subsection{Ti$^{3+}$ reference spectra from stoichiometric LaTiO$_3$ thin films}
{
The complex over-oxidation mechanism in LTO thin films has led to some confusion about the intrinsic line shape of pure Ti$^{3+}$ spectra of LTO in the literature. \cite{disa_orbital_2015,abbate_soft-x-ray-absorption_1991,lin_final-state_2015} Our systematic study of the over-oxidation and its impact on the titanium valency in LTO thin films allows us to fabricate and identify stoichiometric samples. We therefore provide here the Ti 2$p$ photoemission core level spectra and Ti $L$-edge absorption spectra of purely tetra- and trivalent titanates for future reference.
The Ti $2p$ core level line measured at 3$\;$keV photon energy is displayed in Figure \ref{FigS6}(a) together with a Ti$^{4+}$ reference obtained from Niobium-doped SrTiO$_3$. The Ti $L$-edge x-ray absorption spectrum measured in total electron yield of the same samples can be found in Figure \ref{FigS6}(b). The Ti$^{3+}$ spectra are clearly distinguishable from the Ti$^{4+}$ spectra making a straight forward identification of the titanium valency possible. The chemical shift between the two compounds detected in the Ti $2p$ spectrum is about $1.1\;$eV and the full width at half maximum of the Ti$^{3+}$ is significantly increased compared to the Ti$^{4+}$ spectrum. The broadening arises from the interactions of the Ti $2p$ core hole and the Ti $3d$ electron, which is absent for the Ti$^{4+}$ case. The Ti$^{3+}$ x-ray absorption spectrum exhibits a rich multiplet structure resulting in broader features compared to the Ti$^{4+}$ spectrum.

\begin{figure*}[hbpt]
{
\includegraphics[width = 0.99\textwidth]{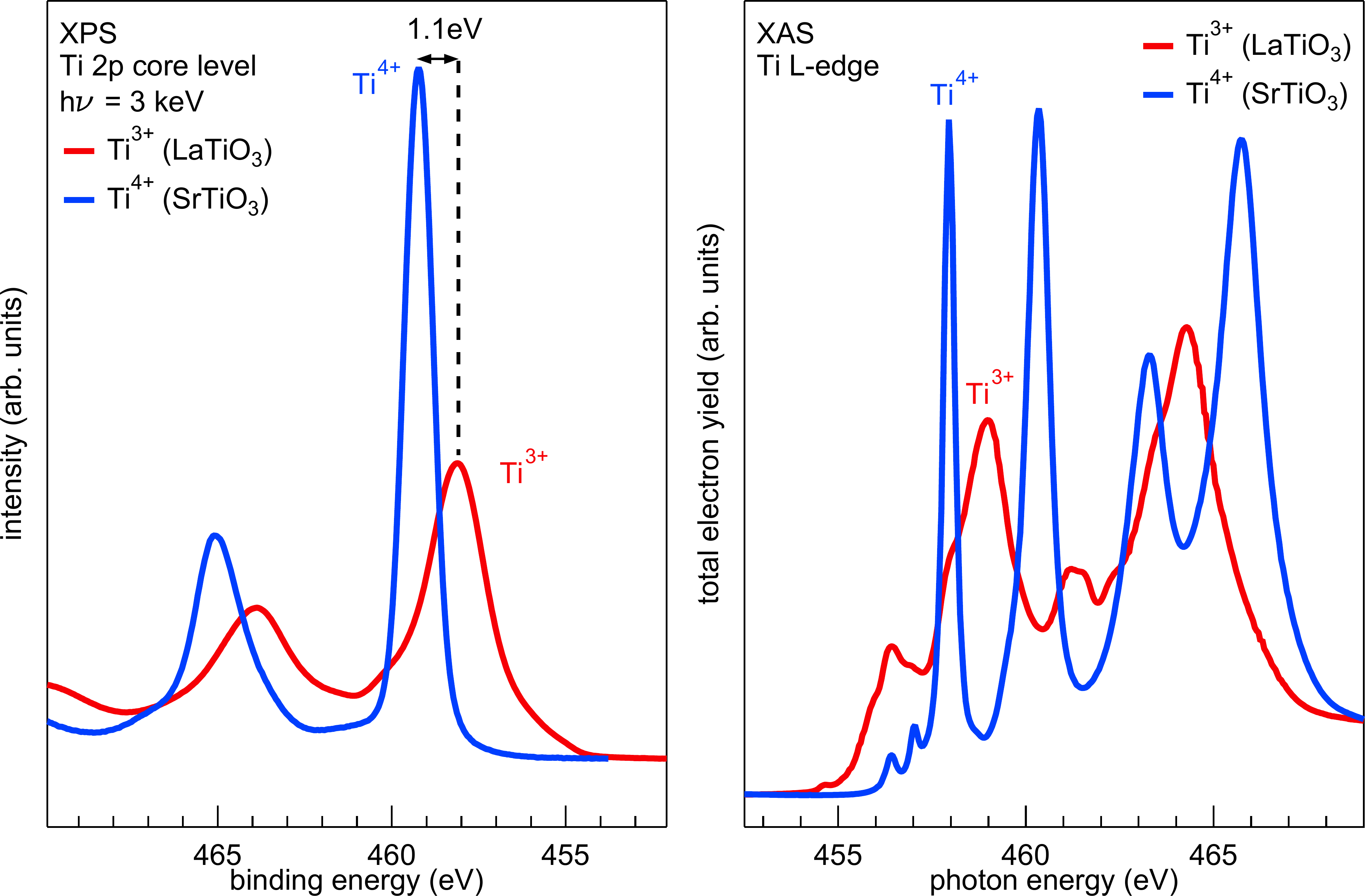}
}
\caption{
\label{FigS6}
Ti $2p$ photoemission (a) and Ti $L$-edge x-ray absorption (b) spectra of stoichiometric LaTi$^{3+}$O$_3$ thin films plotted together with spectra measured on SrTi$^{4+}$O$_3$. The spectra are normalized to equal integral spectral weight of the Ti $2p$ and the Ti $L$-edge for photoemission and x-ray absorption data, respectively.
}
\end{figure*}
}

%

\bibliographystyle{apsrev4-1}
%